\documentclass[floats,floatfix,showpacs,amssymb,prd,twocolumn,superscriptaddress,nofootinbib,nolongbibliography,reprint]{revtex4-1}

\usepackage{amssymb,amsmath,verbatim,mathtools,needspace,enumitem,etoolbox,graphicx,physics,microtype,afterpage,xspace,tabularx,lmodern,multirow}
\usepackage{gensymb}
\usepackage[dvipsnames, usenames]{xcolor}
\definecolor{linkcolor}{rgb}{0.0,0.3,0.5}
\usepackage[unicode, colorlinks=true, linkcolor=linkcolor, citecolor=linkcolor, filecolor=linkcolor, urlcolor=linkcolor, linktocpage, breaklinks]{hyperref}
\usepackage[all]{hypcap}
\usepackage[T1]{fontenc}
\usepackage[utf8]{inputenc}
\usepackage[usenames,dvipsnames]{xcolor}
\hypersetup{colorlinks=true,citecolor=romared,linkcolor=romared,urlcolor=romared}

\definecolor{romared}{RGB}{142,0,28}

\newcommand{\be}{\begin{equation}}
\newcommand{\ee}{\end{equation}}

\def\be{\begin{equation}}
\def\ee{\end{equation}}
\newcommand{\beq}{\begin{eqnarray}}
\newcommand{\eeq}{\end{eqnarray}}
\usepackage{makecell}
\usepackage{soul}
\usepackage[nolist,nohyperlinks]{acronym}

\def\Lie{\mathcal{L}}
\def\p{\partial}
\def\dif{{\rm{d}}}

\def\mV{m_{\rm V}}

\begin{document}

\title{The problem with Proca: ghost instabilities in self-interacting vector fields}

\author{Katy Clough}
\email{k.clough@qmul.ac.uk}
\affiliation{School of Mathematical Sciences, Queen Mary University of London Mile End Road, London, E1 4NS, UK}
\author{Thomas Helfer}
\email{thelfer1@jhu.edu}
\affiliation{Department of Physics and Astronomy, Johns Hopkins University,
3400 N. Charles Street, Baltimore, Maryland, 21218, USA}
\author{Helvi Witek}\email{hwitek@illinois.edu}
\affiliation{Illinois Center for Advanced Studies of the Universe, Department of Physics, University of Illinois at Urbana-Champaign, Urbana, IL 61801, USA}
\date{Received \today; published -- 00, 0000}
\author{Emanuele Berti}
\email{berti@jhu.edu}
\affiliation{Department of Physics and Astronomy, Johns Hopkins University,
3400 N. Charles Street, Baltimore, Maryland, 21218, USA}

\begin{abstract}
Massive vector fields feature in several areas of particle physics, e.g., as carriers of weak interactions, dark matter candidates, or as an effective description of photons in a plasma. Here we investigate vector fields with self-interactions by replacing the mass term in the Proca equation with a general potential. We show that this seemingly benign modification inevitably introduces ghost instabilities of the same kind as those recently identified for vector-tensor theories of modified gravity (but in this simpler, minimally coupled theory). It has been suggested that nonperturbative dynamics may drive systems away from such instabilities. We demonstrate that this is not the case by evolving a self-interacting Proca field on a Kerr background, where it grows due to the superradiant instability. The system initially evolves as in the massive case, but instabilities are triggered in a finite time once the self-interaction becomes significant. These instabilities have implications for the formation of condensates of massive, self-interacting vector bosons, the possibility of spin-one bosenovae, vector dark matter models, and effective models for interacting photons in a plasma.
\end{abstract}

\maketitle

\noindent {\bf \em Introduction.}
Massive vector fields provide a toy model for massless gauge bosons, e.g. photons, that acquire an effective mass in a plasma, and Beyond Standard Model (BSM) ``dark photons,'' which are a candidate for dark matter~\cite{Holdom:1985ag,Agrawal:2018vin,Jaeckel:2010ni,Goodsell:2009xc}.  The Proca equation~\cite{Proca:1900nv} describes the evolution of a massive vector field in the classical limit.
Massive vector particles (the W and Z bosons) also exist in the Standard Model, obtaining their mass via the Higgs mechanism, but they decay rapidly and so do not form condensates. 

Self-interactions arise naturally in extensions of the Standard Model, or in effective descriptions of massive vector interactions with other particles~\cite{Freitas:2021cfi,Fukuda:2019ewf, Hell:2021wzm}. In the photon plasma case, for example, they arise from the induced four-photon interaction in the low-energy limit of QED~\cite{2006physics...5038H}, or from electromagnetic fields of the photon cloud interacting with the surrounding plasma~\cite{Conlon:2017hhi}.
It therefore seems natural to add, as a toy model for such interactions, terms in the Lagrangian that are composed of higher-order products of the field.
Self-interacting vector fields have also been studied in the context of exotic compact objects, in particular ``Proca stars''~\cite{Liebling:2012fv,Brito:2015pxa,Brihaye:2017inn,Sanchis-Gual:2017bhw,Sanchis-Gual:2018oui,DiGiovanni:2018bvo,Zhang:2021xxa,Jain:2021pnk,Herdeiro:2021lwl,Amin:2022pzv,Gorghetto:2022sue}, which may be partly stabilized by a quartic potential~\cite{Minamitsuji:2018kof,Herdeiro:2020jzx}.

Recent work on vectorization, i.e., the  spontaneous growth of vector fields in vector-tensor modified gravity models has uncovered ghost type instabilities which occur when there is a change in the signature of the effective metric for the scalar mode of the Proca field~\cite{Silva:2021jya,Garcia-Saenz:2021uyv,Demirboga:2021nrc,Oliveira:2020dru}.
In such theories, the instability is triggered by non-minimally coupled vector perturbations, and is closely related to the presence of a potential tachyonic mode.

In this work we show that the same ghost instabilities exist in the case of the vector potential that we motivated above, and are therefore not purely a consequence of non-minimal coupling to general relativity (GR).
In particular, we show that they exist for massive vector fields with self-interaction potentials within Einstein's gravity.
We first derive the instability, following the approach of Ref.~\cite{Silva:2021jya}, and then illustrate that the instability can be reached dynamically, something that could not be shown by the analysis of Refs.~\cite{Silva:2021jya,Garcia-Saenz:2021uyv,Demirboga:2021nrc}. We demonstrate this by simulating the superradiant growth of a Proca field around a Kerr black hole (BH) until self-interactions become important.

The rapid superradiant growth of Proca fields around Kerr BHs has been studied using analytical and numerical techniques~\cite{Dolan:2018dqv, Cardoso:2018tly, Pani:2012vp, Pani:2012bp, Rosa:2011my, Endlich:2016jgc,Baryakhtar:2017ngi,Frolov:2018ezx,Baumann:2019eav,Baumann:2019ztm,Tsukada:2020lgt,Santos:2020pmh,Siemonsen:2019ebd,Witek:2012tr,Zilhao:2015tya,East:2017ovw, East:2017mrj}.
Superradiant growth occurs when the Compton wavelength of the vector is comparable to the BH horizon size. Energy and angular momentum are extracted from repeated scattering in the ergosphere, leading to the build-up of the field amplitude~\cite{Brito:2015oca}.
Observations of BH masses and spins can provide constraints on the existence of such light bosonic particles in nature~\cite{Arvanitaki:2009fg,Arvanitaki:2010sy}
(see~\cite{Baryakhtar:2017ngi,Tsukada:2020lgt,Caputo:2021efm,Cardoso:2018tly} for vector constraints).

Due to the faster growth rate of vector bosons compared to the scalar case, fully nonlinear simulations of superradiance with Proca fields can be performed numerically~\cite{East:2017mrj, East:2017ovw}, and have been used as a proxy for the spin-0 case.
Self-interactions in superradiance are of particular interest because they may give rise to observable explosive phenomena, so-called ``bosenovae''~\cite{Yoshino:2012kn, Yoshino:2013ofa, Yoshino:2015nsa,Arvanitaki:2010sy, Sun:2019mqb, Chen:2019fsq,Freitas:2021cfi,Fukuda:2019ewf},
resulting in a collapse of the cloud at a critical point. 
Since the self-interactions of BSM fields are largely unconstrained, the question of how robust superradiance is to them is important for the derived constraints and observables.
For photons confined in a plasma, where an effective mass arises from environmental effects, self-interactions are an essential component of the model ~\cite{Conlon:2017hhi,Cardoso:2020nst,Dima:2020rzg,Blas:2020kaa,Cannizzaro:2021zbp,Cannizzaro:2020uap,Fukuda:2019ewf,Wang:2022hra}.
It has been suggested that self-interactions may quench the superradiant growth before an explosion is reached, but the existing literature only applies to the perturbative regime where $M\mu\ll1$~\cite{Baryakhtar:2020gao,Omiya:2020vji} (we employ geometrical units $G=c=1$). 
One may have hoped to settle the argument using fully nonlinear simulations of a self-interacting vector bosenova. The ghost instabilities we identify appear to make this unviable, since they occur precisely when the nonlinear attractive terms begin to dominate. 
Superradiance around a BH may seem like a special case in which to test the ghost instability identified here. However the ghost does not depend on a particular metric, and could be triggered in other cases where the field has a source that causes it to grow in amplitude.
The advantage of studying the superradiant system is that the BH slowly feeds energy into the vector field, which enables us to follow the system from the massive regime, where no ghost instability exists, into the nonlinear regime where self-interactions become relevant, and demonstrate than the evolution can reach the instability point. In addition, while the Kerr background metric is axisymmetric, the superradiant field profile itself is not, and thus one cannot claim that symmetries in the field impose any particular restrictions on the system.

This work is organized as follows.
We first give the analytical justification for the instability, following Ref.~\cite{Silva:2021jya}. We show that there are in fact two potential instabilities: one related to the change in sign of the effective metric, and one of tachyonic nature.
We then describe our numerical simulations, that demonstrate explicitly that these instabilities are triggered in the case of a growing vector cloud around a BH. Finally we discuss our findings and their implications.

The Supplemental Material contains more detailed steps of our analytic calculation, the full ADM decomposition of the Proca equations of motion with self interactions used in our simulations, numerical testing and convergence results. This material includes Refs. \cite{Silva:2021jya,Demirboga:2021nrc,East:2017ovw, East:2017mrj, Witek:2012tr,Visser:2007fj,Palenzuela:2009hx,Hilditch:2013sba,Zilhao:2015tya,Dolan:2018dqv,Clough:2015sqa,Andrade2021,Radia:2021smk}.

\noindent {\bf \em Analytical evidence for the instability.}
Here we provide analytic evidence for the presence of ghost and tachyonic instabilities in self-interacting vector fields (see the Supplementary Material for more details).

Consider a massive vector field $X^{\mu}$ with mass $\mu$, norm $X^2 = X^\mu X_\mu$ described by the Lagrangian
\begin{align}
\label{eq:LagrangianProcaSI}
  \Lie = & \frac{1}{16\pi}\,R - \frac{1}{4} F^{\mu\nu} F_{\mu\nu} 
        - V\left(X^2 \right)
\,,
\end{align}
where $R$ denotes the four-dimensional Ricci scalar, the Maxwell tensor is
$F_{\mu\nu} = \nabla_{\mu}X_{\nu} - \nabla_{\nu} X_{\mu}$, $V\left(X^2 \right)$ is a generic potential.
The vector field equation is then
\begin{equation}
\label{eq:GenVEoM}
    0 = \nabla^{\nu} F_{\mu\nu} + \frac{\dif V}{\dif X^{\mu} }
    = \nabla^{\nu} F_{\mu\nu} + \mu^2 z X_{\mu} 
\,,
\end{equation}
where $\frac{\dif V}{\dif X^{\mu}}= 2 X_{\mu} V' $,
with $V' = d V / d (X^2)$,
and we have introduced the scalar quantity
\begin{equation}
\label{eq:GenVz}
    z = \frac{2}{\mu^2} V' (X^2)
\,.
\end{equation}

One obtains a modified Lorenz condition by taking the derivative of~\eqref{eq:GenVEoM} and using the symmetries of the Maxwell tensor,
\begin{equation}
\label{eq:GenVLG}
    \nabla^{\mu}\left( z X_{\mu} \right) =0 ~.
\end{equation}
To identify the conditions on the potential $V(X^2)$ (or on $z$) for which ghost instabilities appear we follow the strategy of Ref.~\cite{Silva:2021jya} and rewrite the field equation~\eqref{eq:GenVEoM} as a wave equation for the vector field $X^{\mu}$ with an effective ``mass'' matrix. Since $z$ depends strongly on $X^\mu$, and we will be interested in cases where the self-interactions are small but not negligible, we need to include additional terms in the principal part to find the instability condition. We can then introduce an effective metric, writing the highest derivatives of $X^{\mu}$ as a wave operator, to determine the presence of the ghost instability.

We start by expanding the field Eq.~\eqref{eq:GenVEoM} in terms of the vector field $X^{\mu}$. Inserting Eq.~\eqref{eq:GenVLG} to replace the divergence of the field, $\nabla_{\mu}X^{\mu}$, yields
\begin{align}
    0 = & \nabla^{\nu} \nabla_{\nu} X_{\mu} - \mu^2 z X_{\mu} - R_{\mu\nu} X^{\nu}
    \nonumber\\ &
    + X^{\nu} \frac{\nabla_{\mu}\nabla_{\nu} z}{z}
    - X^{\nu} \frac{\nabla_{\mu}z \nabla_{\nu}z} {z^2}
    + \nabla_{\mu} X^{\nu} \frac{\nabla_{\nu}z}{z}
    \,.
\end{align}
This equation contains mixed second derivatives of $X^{\mu}$, in addition to the wave operator, that contribute to its principal symbol. Using that
$z' = \frac{2}{\mu^2} V''$,
and rearranging terms, we get the wave equation
\begin{equation}
\label{eq:GenVWavEq}
    \hat{g}_{\rho\sigma} \nabla^{\rho} \nabla^{\sigma} X_{\mu} - \mathcal{M}_{\mu\rho} X^{\rho} = 0 
\,.
\end{equation}
Here
\begin{equation}
\label{eq:GenVgeff}
    \hat{g}_{\mu\nu} = \frac{2}{\mu^2} \left[ V' g_{\mu\nu} + 2 V'' X_{\mu} X_{\nu} \right]
\end{equation}
and we have introduced the matrix
\begin{align}
\label{eq:MassMatrix}
    \mathcal{M}_{\mu\nu} = & z R_{\mu\nu} 
        + z \left(\mu^2 \hat{g}_{\mu\nu} - 3 V'' X_{\mu} X_{\nu}  \right)
        + {\rm l.o.t.}
\end{align}
where ``l.o.t.'' denotes lower order terms $\sim \nabla X^{\mu}$, that do not contribute if we expand the vector field $X^{\mu}=X^{\mu}_{0}+\epsilon \delta X^{\mu}$ around a constant $X^{\mu}_{0}$.
$\mathcal{M}_{\mu\nu}$ can be interpreted as an effective mass matrix for perturbations of the vector field if back-reaction onto the spacetime is negligible.

The ghost instability appears if the effective metric satisfies $\hat{g}^{tt}>0$~\cite{Silva:2021jya,Demirboga:2021nrc}.
We identify the condition for the instability by contracting Eq.~\eqref{eq:GenVgeff} with the time-like vector $n^{\mu}$ that is normal to the spatial hypersurfaces (see Supplementary Materials for further details), decomposing the vector field into the scalar potential $\varphi$ and spatial vector $A^{\mu}$ as $X^{\mu} = A^{\mu} + n^{\mu}\varphi$.
We then find
\begin{equation}
\label{eq:GenVGhostCondition}
\hat{g}_{nn} = \hat{g}_{\mu\nu} n^{\mu} n^{\nu}
    = 2 \varphi^2 z' - z
    = \frac{2}{\mu^2} \left[2 \varphi^2 V'' - V' \right]
\,.
\end{equation}
Note that in these expressions, the effect of curvature appears only in the mass matrix via $R_{\mu\nu}$, and not in $\hat{g}_{nn}$. Therefore, spacetime curvature (including that sourced by other matter) does not play a role in the ghost instability, which could in principle be triggered in flat space for some large amplitude of the field components.

Consider as an illustration the quartic potential
\begin{align}
\label{eq:DefPotentialInX}
V(X^2) = & \frac{\mu^2}{2} X^2
        + \frac{\lambda \mu^2}{4} (X^2)^2
\,,
\end{align}
which includes the mass term of the Proca field
(with mass $\mV=\mu \hbar$)
and a self-interaction potential. We test this case numerically in the following section.
The value of $\lambda$ is determined by the strength and sign of the self-interaction arising from higher energy physics: naively, a positive (negative) $\lambda$ corresponds to a repulsive (attractive) self-interaction, but since $X^2$ is not positive definite, the actual nature of the self-interaction will vary depending on the local value of the vector field.
As explained further in the Discussion, the value of $\lambda$ is related to a fundamental symmetry breaking scale $f_a$ (as $|\lambda| \sim 1/f^2_a$) that one expects to be lower than the Planck mass\footnote{If one assumes that dark photons make up a significant proportion of the dark matter, then self-interactions must be small and the symmetry breaking scale high, due to the constraints from structure formation and the Bullet Cluster observations \cite{Randall:2008ppe}. However, vector superradiance can occur where the dark photons have no coupling to Standard Model matter, and do not make up a significant part of the dark matter, so the parameter $\lambda$ is in principle unconstrained by observations.}.
In the strong superradiance regime we study, this means that our results would be relevant where $|\lambda| \gtrsim 1$. 
The quartic potential yields
$V' = \frac{\mu^2}{2}\left(1 + \lambda X^2 \right)$ and
$V'' = \frac{1}{2}\lambda \mu^2 \neq0$,
and the ghost should appear if
$\hat{g}_{nn} = - 1 - \lambda X^{2} + 2 \lambda \varphi^2 \ge 0$. We find that positive values of $\lambda$ are more susceptible to this instability (see Fig. \ref{fig-SR_timeprofiles}).

We also consider the conditions under which a tachyonic instability arises, i.e., where an eigenvalue of the mass matrix~\eqref{eq:MassMatrix} changes sign, or equivalently when its determinant is equal to zero. In the quartic case, assuming a Minkowski background and keeping only up to first order in $\lambda$, this happens when
$\left(1+\frac{17}{2} X^\mu X_\mu\right)< 0$. We find that this instability is triggered first for negative $\lambda$ (see Fig. \ref{fig-SR_timeprofiles}).

We will demonstrate below that both the ghost and tachyonic instabilities derived here correspond to failures observed in numerical simulations, indicating that such unstable states can be reached dynamically from physically reasonable initial conditions.
Although we illustrate the instabilities using the case of a fourth-order potential, we note that the instabilities are not related to the unboundedness of the potential from below but to the presence of inflection points where the potential changes from convex to concave, and so they cannot be ``cured'' by adding higher-order terms. We have tested this finding numerically by adding sixth-order corrections which are positive definite, and find consistent results.

\noindent {\bf \em Numerical evolution to instability.}
We now perform numerical simulations of the time-domain evolution of a self-interacting Proca field around a Kerr BH of mass $M$ and spin $J=a\,M$ in $3+1$ dimensions, with the quartic potential of Eq.~\eqref{eq:DefPotentialInX}.

Following~\cite{East:2017ovw, East:2017mrj, Witek:2012tr}, we use an ADM decomposition of the metric in Cartesian Kerr-Schild (KS) coordinates (the full metric is given in the Supplementary Material),
considering the case where this is a background metric on which the dynamical Proca field evolves (i.e. we neglect backreaction of the field onto the metric).
As above, we decompose the vector field into a scalar $\varphi$ and a purely spatial vector $A^{\mu}$ and define the electric field as $E_i = -\alpha F_{i0}$ where $\alpha$ is the lapse function of the ADM metric. The evolution equations for these quantities and further details of the numerical scheme are given in the Supplementary Material.

\begin{figure}
\includegraphics[width=8cm]{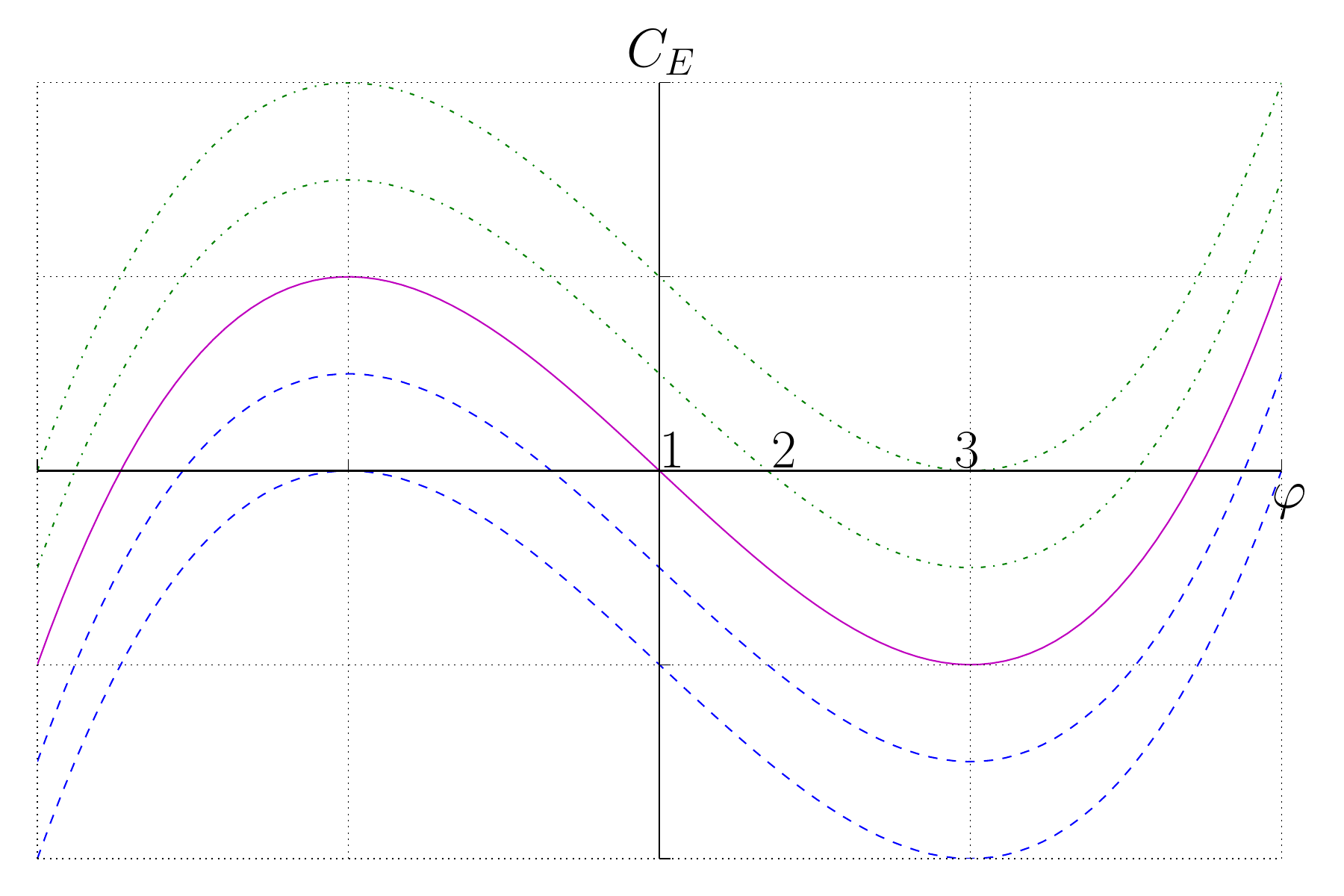}
\caption{
The Proca evolution equation gives rise to a cubic constraint equation for $\varphi$ of the form
$C_{E} = a \varphi^3 + b\varphi + c = 0$,
with $a = -\mu^2\lambda$, $b = \mu^2(1+A_iA^i)$ and $c=D_i E^i$. 
Each curve in this plot depicts the constraint $C_{E}$ at different times at a given point in space, to illustrate how solutions are lost when the ghost instability is triggered.
Solutions exist for $\varphi$ at the intersections with the $x$ axis, where $C_{E}=0$. The solid pink line shows a general point at $t = 0$, when we choose $D_iE^i=0$ and $\varphi=0$ (point 1). The dot-dashed green lines show subsequent events, where the root has moved to the positive values at 2 and then to 3, where $\hat{g}_{nn}=0=dC_{E}/d\varphi = 0$. Beyond this point, as the field continues to grow, there is no continuous real solution to the constraint. One could switch to the negative $\varphi$ solution at this point, but this would create a discontinuity.
Other points in the spacetime evolve over time as shown by the blue dashed lines, for which it is the negative $\varphi$ solution that disappears -- thus there is no obvious way to choose, a priori, which root to follow if not the central one.
\label{fig-cubic}
}
\end{figure}

\begin{figure}[t]
\includegraphics[width=0.48\textwidth]{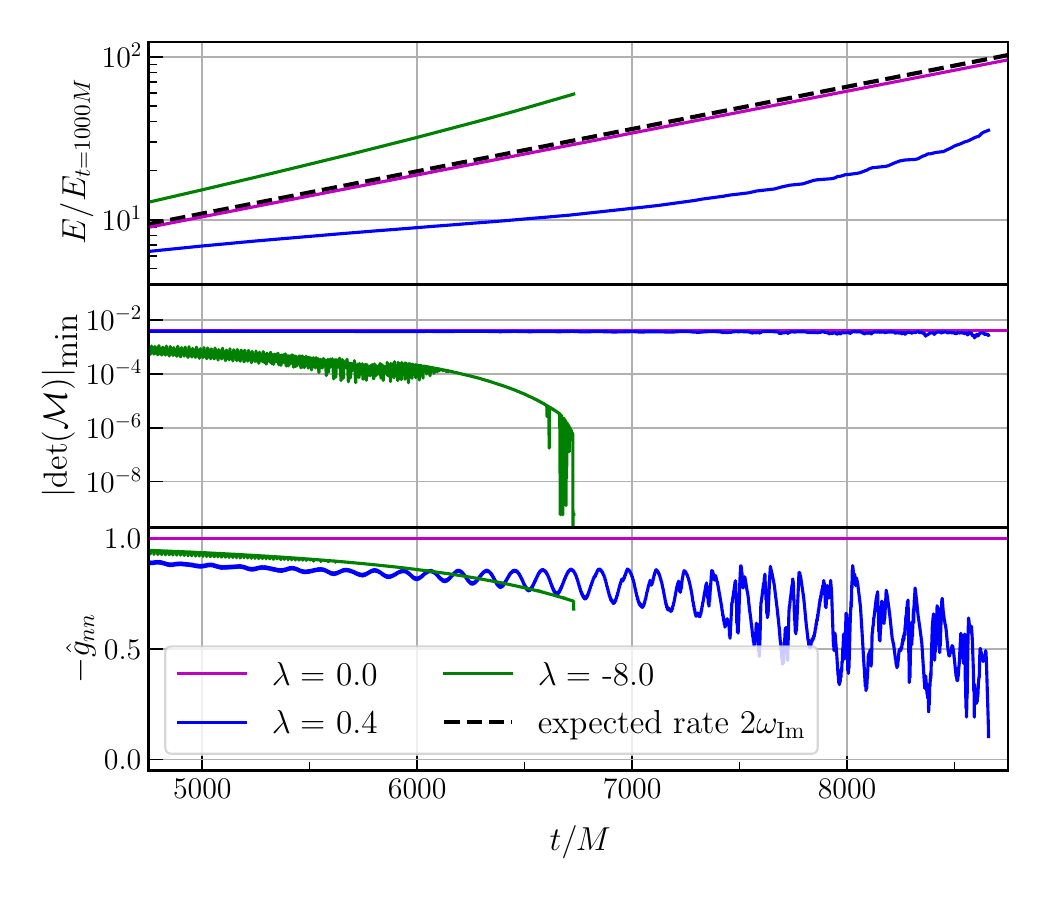}
\caption{
Late-time evolution of the superradiant growth and instabilities for quartic self-interactions $\lambda = 0, -8.0, 0.4$. Top panel: growth of the cloud's total energy (normalized by the energy at $t/M=1000$).
Positive self-interactions suppress the growth relative to the purely massive case, whereas negative ones enhance it.
Middle panel: evolution of the determinant of the matrix $\mathcal{M}$, showing that it goes to zero
and triggers the tachyonic instability for $\lambda=-8.0$ around $t/M \sim 6750$. Bottom panel: $\hat{g}_{nn}$ goes to zero at the onset of the ghost instability.
For $\lambda=0.4$ it is triggered at $t/M \sim 8650$. 
\label{fig-SR_timeprofiles}
}
\end{figure}

A key feature of the decomposition is that we obtain an evolution equation for $\varphi$ of the form
\begin{multline}
    \dif_{t} \varphi = - A^{k} D_{k}\alpha
        - \frac{\alpha}{\hat{g}_{nn}} \left(1 + \lambda X^2 \right) \left[ \varphi\,K - D_{k} A^{k} - Z\right] \\
        + \frac{2 \lambda \alpha }{\hat{g}_{nn}} \left[ A^{i} A^{j} D_{i} A_{j} - \varphi \left( E^{k} A_{k} - K_{ij} A^{i} A^{j} + 2 A^{k} D_{k}\varphi \right)
         \right]
\,,
\end{multline}
where $K_{ij}$ and $K=\gamma^{ij}K_{ij}$ are the extrinsic curvature and its trace,
$D_{i}$ is the covariant metric related to the spatial metric, $Z$ is an auxiliary vector, $X^2 = X^\mu X_\mu = A^i A_i - \varphi^2$ and $\dif_{t}$ is the advective derivative in the normal direction. The quantity $\hat{g}_{nn}$ appears in the condition for the ghost instability of Eq.~\eqref{eq:GenVGhostCondition}.
It is clear that divergences may arise if $\hat{g}_{nn}\rightarrow0$, but it is not immediately obvious that this is fatal, since the quantities in the numerator may tend to zero at the same rate.
However, if we consider the constraint in the 3+1 form,
\begin{align}
C_{E} = & D_{i} E^{i} + \mu^2 \varphi \left(1 + \lambda X^2 \right)
        = 0
\,,
\end{align}
and note that $dC_{E}/d\varphi = \hat{g}_{nn}$
it becomes clear
that $\hat{g}_{nn} =0$ marks the point at which the simulation cannot continue. As illustrated in Fig.~\ref{fig-cubic}, the condition $\hat{g}_{nn}=0$ corresponds to the point at which the constraint equation (a cubic equation) loses two of its roots. Once $\hat{g}_{nn}=0$ at any point in the spacetime, there are no longer continuous real solutions to the constraint, so it becomes impossible to continue the evolution any further. This observation provides another interpretation of the ghost instability identified in this work, and corresponds to the behavior we observe in the simulations.
A similar behavior may be expected for higher-order potentials, which lead to constraint equations of higher-than-cubic order.

%
A summary of the late-time evolutions for $A_0 = 0.1$, $\mu M= 0.5$, dimensionless spin $a/M = 0.99$, and three values for the self-interaction $\lambda = 0, -8.0, 0.4$ are shown in Fig.~\ref{fig-SR_timeprofiles}.
We find that, following an initial transient period, the field energy in all cases grows smoothly at the expected superradiant rate obtained in~\cite{East:2017mrj, Dolan:2018dqv}.
However, eventually self-interactions become important and affect the growth rates.
Instabilities occur at the point at which either $\det(\mathcal{M}_{\mu\nu})=0$ 
(tachyonic instability; for negative $\lambda$)
or $\hat{g}_{nn}\rightarrow0$
(ghost instability; for positive $\lambda$).
When the tachyonic instability is excited, the field amplitude grows rapidly, which will ultimately also trigger the ghost. Up until the instabilities occur the solutions are regular and show no sign of blow-up.
To get closer to the instability points we need to significantly reduce our timestep size, indicating that the dynamical timescales are decreasing, but there is no evidence of any dynamical mechanism to avoid the instability.

Our simulations do not account for backreaction, but this does not affect our main conclusions. There is a scaling freedom in the value of $\lambda$ and the initial field amplitude $A_0$, with the quantity $\lambda (A_0)^2 \sim 1$ triggering the instabilities. Thus the simulations accurately represent a regime early in the superradiant build up where the field amplitude is sufficiently small that backreaction on the metric is indeed negligible, at which point one will trigger an instability via self-interactions provided $\lambda$ is sufficiently large. The results can also be scaled to higher amplitudes representing later points in the growth, at which point a smaller self-interaction strength would trigger the instability, but in this case the assumption of small backreaction is no longer as accurate and one should account for the spin-down of the BH. This approximation will be considered further in the Discussion, where we find that, for the fastest growing modes, self-interactions corresponding to a fundamental scale smaller than the Planck mass should always become relevant before the superradiant growth saturates.
Finally, we note that the local curvature does not play an important role in the ghost instability. Thus there is no reason for superradiance to be an exceptional example in which the ghost is triggered.

\noindent {\bf \em Discussion.}
Recent works have shown that modified vector-tensor theories possess ghost instabilities, calling into question the viability of spontaneous vectorization of compact objects. Here we have shown that similar instabilities affect minimally coupled vector fields in general relativity, when the field possesses a non trivial self-interaction potential. 

We illustrated the ghost and tachyonic instabilities for a fourth-order potential, but note that these may occur for any potential that contains an inflection point (and thus a tachyonic region). We related the ghost instability to the fact that the vector field ``sees'' an effective metric with a signature that changes sign, as in Refs.~\cite{Silva:2021jya,Garcia-Saenz:2021uyv,Demirboga:2021nrc}. We also related it to the fact that continuous solutions to the constraint equation cease to exist. 

It has been suggested that in a fully nonlinear dynamical evolution, and in the absence of symmetries in the field, the system may somehow avoid such unstable solutions. We demonstrate that this is not the case by evolving the fully self-interacting, nonaxisymmetric equations of superradiant growth from the massive Proca regime to the point when the self-interactions become important. The evolution continues right up until the instability criterion is met at a single event, after which the simulation breaks down. Whilst the specific timing of breakdown will be related to the formulation of the problem, in particular the chosen coordinates of the evolution, it seems likely that the instability would be reached at some point regardless of the chosen coordinates once the field amplitude is large. It can occur even in Minkowski spacetimes, where there are fewer coordinate ambiguities, thus it seems to be an indication of a fundamental pathology of the self-interacting Proca equations, and not a problem of the numerical formulation. However, further work is required to elucidate this point.
Our simulations are in the decoupling limit, i.e., they neglect gravitational backreaction, but this is a very good approximation when the density of the cloud is low, as it will be during most of the superradiant build-up.

We can estimate a physical value for the self-interaction strength at which ghosts are relevant. Our numerical results possess a scaling symmetry, as they depend only on the (dimensionless) combination $\lambda X^2$.
The instabilities should arise when $\lambda X^2\sim 1$, but looking at the maximum values obtained in our simulations before they break down we have observed that they can already be problematic for values as small as $\lambda X^2 \sim 0.01$. If we consider the effect of backreaction, and the resulting saturation of the superradiant instability at some cloud mass $M_{\rm cloud}^{\rm max}$, this fixes the maximum field amplitude for a given $\mu M$, and we can then relate the self-interaction strength to an energy scale of the vector potential $f_a$, where $\lambda \sim 1/f_a^2$. Considering the superradiant build-up without any self-interaction, the amplitude of the field at saturation of the instability can be estimated from the maximum cloud mass using the fact that for a superradiant cloud
$M_{\rm cloud}/M \sim X^2/(\mu M)$.
Therefore self-interactions will be relevant when
\begin{equation}
    f_a \lesssim \sqrt{\mu M \left( \frac{M_{\rm cloud}^{\rm max}}{0.01 M} \right)} ~ {\rm M_{\rm Planck}} \label{eq-bound}
\end{equation}
in Planck units. For superradiance to occur one requires that $\mu M \sim 1$, and cloud masses of order 1\% of the BH mass are typically generated for the fastest growing modes. For example, in the case studied $\mu M = 0.5$, and $M_{\rm cloud}^{\rm max} \sim 0.02M$ at saturation (see Fig.~1 of \cite{East:2017ovw}). Thus Eq.~\eqref{eq-bound} indicates that instabilities should develop before saturation of the cloud in any case of vector superradiance in which self-interactions manifest below the Planck scale. In the presence of backreaction one may obtain a slightly weaker result because the spin-down would slow the cloud growth in the later stages, and mode mixing could result in the cloud saturating at a smaller mass. However, this estimate is likely to hold as an order-of-magnitude approximation.

The key cause of the ghost instability appears to be the loss of the longitudinal degree of freedom when the effective mass becomes (instantaneously, and locally at any event) zero~\cite{Kribs:2022gri}. This degree of freedom decouples from the Lagrangian and, even though a mass is recovered, it no longer possesses a continuous solution. The result is that vector potentials with non-convex regions are fundamentally unstable. Another interpretation, coming from the effective field theory approach~\cite{deRham:2018qqo}, is that for the effective mass to vanish, one requires $\mu^2 \lambda X^\mu X_\mu \sim \Lambda^2$, where $\Lambda$ is a cut off above which the theory breaks down. However, this interpretation is based on Wilsonian renormalization flows, which are not reflected in the classical treatment. Nevertheless, it provides an insight into how vector self-interactions based on the Higgs mechanism or Euler-Heisenberg Lagrangian may avoid the ghost.

This work has implications for a broad range of physical scenarios, including studies of self-interacting vector fields in ``bosenova'' type events, both as an effective model for photons in a plasma and for dark photons in BSM models. The instability found here calls into question the validity of such studies beyond the purely massive case (where these instabilities do not occur). 

\noindent {\bf \em Note added in proof.}
Since our work appeared, others have highlighted similar issues ~\cite{Coates:2022qia,Aoki:2022woy,Barausse:2022rvg,Mou:2022hqb}, with a preprint of \cite{Mou:2022hqb} appearing on the same day as ours. In particular our result has been linked to the idea that the model of mass plus mass plus polynomial corrections breaks down as an effective theory, and one needs to employ a more complete theory with additional fields (such as an Abelian Higgs mechanism, as in Ref.~\cite{East:2022ppo}) to provide the mass once the instability scales are reached~\cite{Coates:2022qia,Aoki:2022woy,Barausse:2022rvg}.

\acknowledgements

\noindent We thank Jean Alexandre, Mustafa Amin, Andrea Caputo, Alexandru Dima, Pedro Ferreira, Mudit Jain, Yoni Khan, Scott Melville, Zong-Gang Mou, Jens Niemeyer, Hector Okada da Silva, Surjeet Rajendran and Hong-Yi Zhang for helpful discussions.
KC acknowledges funding from the European Research Council (ERC) under the European Union’s Horizon 2020 research and innovation programme (grant agreement No 693024), and an Ernest Rutherford Fellowship from UKRI (grant number ST/V003240/1).
T.H. and E.B. are supported by NSF Grants No. PHY-1912550, AST-2006538, PHY-090003 and PHY-20043, and NASA Grants No. 19-ATP19-0051 and 20-LPS20-0011. 
H.~W. acknowledges support provided by NSF grants 
No. OAC-2004879 and No. PHY-2110416, and Royal Society (UK) Research Grant RGF\textbackslash R1\textbackslash 180073.

This research project was conducted using computational resources at the Maryland Advanced Research Computing Center (MARCC).
The authors acknowledge the Texas Advanced Computing Center (TACC) at The
University of Texas at Austin for providing HPC resources that have contributed
to the research results reported within this paper. URL:
http://www.tacc.utexas.edu~\cite{10.1145/3311790.3396656}.
The simulations presented in this paper also used the Glamdring cluster, Astrophysics, Oxford, DiRAC resources under the projects ACSP218 and ACTP238 and PRACE resources under Grant Numbers 2020225359 and 2018194669. This work was performed using the Cambridge Service for Data Driven Discovery (CSD3), part of which is operated by the University of Cambridge Research Computing on behalf of the STFC DiRAC HPC Facility (www.dirac.ac.uk). The DiRAC component of CSD3 was funded by BEIS capital funding via STFC capital grants ST/P002307/1 and ST/R002452/1 and STFC operations grant ST/R00689X/1. In addition used the DiRAC at Durham facility managed by the Institute for Computational Cosmology on behalf of the STFC DiRAC HPC Facility (www.dirac.ac.uk). The equipment was funded by BEIS capital funding via STFC capital grants ST/P002293/1 and ST/R002371/1, Durham University and STFC operations grant ST/R000832/1. DiRAC is part of the National e-Infrastructure. The PRACE resources used were the GCS Supercomputer JUWELS at J\"ulich Supercomputing Centre(JCS) through the John von Neumann Institute for Computing (NIC), funded by the Gauss Centre for Supercomputing e.V. (\url{www.gauss-centre.eu}) and computer resources at SuperMUCNG, with technical support provided by the Leibniz Supercomputing Center.
This work used the Extreme Science and Engineering Discovery Environment (XSEDE) Expanse through the allocation TG-PHY210114, which is supported by NSF Grant No. ACI-1548562.
This work used the Blue Waters sustained-petascale computing project which was supported by NSF Award No. OCI-0725070 and No. ACI-1238993, the State of Illinois and the National Geospatial Intelligence Agency.

\appendix

\section{Supplemental material: Detailed analytic calculation and comparison to prior work}
\label{sec-appendix_analytic}

As in the main text, consider a massive vector field $X^{\mu}$ with norm $X^2 = X^\mu X_\mu$ described by the Lagrangian
\begin{align}
\label{eq:LagrangianProcaSIA}
  \Lie = & \frac{1}{16\pi}\,R - \frac{1}{4} F^{\mu\nu} F_{\mu\nu} 
        - V\left(X^2 \right)
\,.
\end{align}
where $R$ denotes the four-dimensional Ricci scalar, the Maxwell tensor is
$F_{\mu\nu} = \nabla_{\mu}X_{\nu} - \nabla_{\nu} X_{\mu}$, $V\left(X^2 \right)$ is a generic potential, and
we employ geometrical units ($G=c=1$).
The field equation for the vector field is then
\begin{equation}
\label{eq:GenVEoMA}
    0 = \nabla^{\nu} F_{\mu\nu} + \frac{\dif V}{\dif X^{\mu} }
    = \nabla^{\nu} F_{\mu\nu} + \mu^2 z X_{\mu} 
\,,
\end{equation}
where $\frac{\dif V}{\dif X^{\mu}}= 2 X_{\mu} V' $,
with $V' = d V / d (X^2)$,
and we have introduced the scalar quantity
\begin{equation}
\label{eq:GenVzA}
    z = \frac{2}{\mu^2} V' (X^2)
\,.
\end{equation}
In Ref.~\cite{Silva:2021jya}, the authors illustrated how in their modified vector-tensor model, the modified Lorenz condition
\begin{equation}
    \nabla^\mu (\hat{z} X_\mu) = 0 ~
\end{equation}
sets the true effective mass for the field, once it is cast in a manifestly hyperbolic form, as
\begin{equation}
\label{eq-silvaA}
	\nabla_\mu \nabla^\mu X_\alpha + (\nabla_\mu \ln |\hat{z}|)\nabla_\alpha X^\mu =  \mathcal{M}_{\alpha \mu} X^\mu ~.
\end{equation}
In their model the quantity $\hat{z}$ plays a similar role in the instability to our quantity $z$ defined above. However, it depends primarily on the stress-energy tensor of the compact object and the dependence on the vector field is subdominant, so that they obtain the effective mass matrix
\begin{equation}
	\mathcal{M}_{\alpha \beta} = \hat{z} \mu^2 g_{\alpha \beta} + R_{\alpha \beta} - \nabla_{\alpha} \nabla_\beta (\ln{\hat{z}}) ~.
\end{equation}
We can see that when $\hat{z}=0$, there are divergences in the mass term. Viewed another way, the principal part $\hat{z} g^{\mu\nu}\nabla_\mu \nabla_\nu X_\alpha$ is such that the effective metric signature changes sign for $\hat{z}=0$, and thus one develops the negative kinetic terms that are associated with ghosts.

In the case of the self-interacting vector field, one also obtains a modified Lorenz condition by taking the derivative of~\eqref{eq:GenVEoMA} and using the symmetries of the Maxwell tensor,
\begin{equation}
\label{eq:GenVLGA}
    \nabla^{\mu}\left( z X_{\mu} \right) =0 ~.
\end{equation}
To identify the conditions on the potential $V(X^2)$ 
(or, equivalently, on $z$)
for which ghost instabilities appear we
follow the strategy of Ref.~\cite{Silva:2021jya} and rewrite the field equation~\eqref{eq:GenVEoMA} as a wave equation for the vector field $X^{\mu}$ with an effective ``mass'' matrix. Since our $z$ depends strongly on $X^\mu$, and we will be interested in cases where the self-interactions are small but not negligible, we need to include additional terms in the principal part to find the instability condition. We can then introduce an effective metric, writing the highest derivatives of $X^{\mu}$ as a wave operator, which will determine the presence of the ghost instability.

We start by expanding the field equation~\eqref{eq:GenVEoMA} in terms of the vector field $X^{\mu}$. Inserting Eq.~\eqref{eq:GenVLGA} to replace the divergence of the field, $\nabla_{\mu}X^{\mu}$, yields
\begin{align}
    0 = & \nabla^{\nu} \nabla_{\nu} X_{\mu} - \mu^2 z X_{\mu} - R_{\mu\nu} X^{\nu}
    \nonumber\\ &
    + X^{\nu} \frac{\nabla_{\mu}\nabla_{\nu} z}{z}
    - X^{\nu} \frac{\nabla_{\mu}z \nabla_{\nu}z} {z^2}
    + \nabla_{\mu} X^{\nu} \frac{\nabla_{\nu}z}{z}
    \,.
\end{align}
This equation contains mixed second derivatives of $X^{\mu}$, in addition to the wave operator, that contribute to its principal symbol. Using the relation
\begin{equation}
\label{eq:GenVDzA}
    z' = \frac{2}{\mu^2} V''
\,,
\end{equation}
where $'$ again denotes differentiation with respect to $(X^2)$, we obtain

\begin{align}
\label{eq:GenVCalc1A}
    0 = & \nabla^{\nu} \nabla_{\nu} X_{\mu} 
        - \mu^2 z X_{\mu} - R_{\mu\nu} X^{\nu}
        + \frac{4 V''}{\mu^2 z} X^{\nu} X^{\rho} \nabla_{\rho} \nabla_{\mu} X_{\nu}
    \nonumber \\ &
        + \frac{4 V'' X^{\nu} }{\mu^2 z} \nabla_{\mu} X^{\rho} \left(\nabla_{\nu} X_{\rho} + \nabla_{\rho} X_{\nu} \right)
    \nonumber \\ &
        - \left( \frac{16 (V'')^2}{\mu^4 z^2} 
            - \frac{8 V'''}{\mu^2 z}
        \right)
        X^{\nu} X^{\rho} X^{\sigma} \nabla_{\mu} X_{\nu} \nabla_{\rho} X_{\sigma}
\,.
\end{align}
In the next series of steps we will 
(i) reinsert the Maxwell tensor $F_{\mu\nu}=\nabla_{\mu}X_{\nu}-\nabla_{\nu}X_{\mu}$,
and 
(ii) use the field equation~\eqref{eq:GenVEoMA} to replace terms of the form $\sim \nabla_{\rho}F_{\mu\nu}$. If we rewrite Eq.~\eqref{eq:GenVEoMA} as
\begin{equation}
\label{eq:GenVDFmnA}
    0 = g^{\nu\rho} \left[
    \nabla_{\rho} F_{\mu\nu} + \frac{1}{4} \mu^2 z g_{\nu\rho} X_{\mu} 
    \right]
\end{equation}
it follows that the square bracket must vanish up to terms, denoted collectively as $\mathcal{T}_{\mu\nu\rho}$, 
for which $g^{\nu\rho} \mathcal{T}_{\mu\nu\rho}=0$ (which we neglect). 
This allows us to write Eq.~\eqref{eq:GenVCalc1A} as
\begin{align}
\label{eq:GenVCalc2A}
    0 = & \nabla^{\nu} \nabla_{\nu} X_{\mu}
        + \frac{4 V''}{\mu^2 z} X^{\nu} X^{\rho} \nabla_{\nu} \nabla_{\rho} X_{\mu}
        - \mu^2 z X_{\mu} - R_{\mu\nu} X^{\nu}
    \nonumber\\ &
        - V'' \, X^{\nu} X_{\nu} X_{\mu}
        + \frac{4 V'' X^{\nu} }{\mu^2 z} \nabla_{\mu} X^{\rho} \left( \nabla_{\rho} X_{\nu} + \nabla_{\nu} X_{\rho} \right)
    \nonumber\\ &
        - \left( \frac{16 (V'')^2}{\mu^4 z^2} 
            - \frac{8 V'''}{\mu^2 z}
        \right)
        X^{\nu} X^{\rho} X^{\sigma} \nabla_{\mu} X_{\nu} \nabla_{\rho} X_{\sigma}
\,.
\end{align}
We can infer the effective metric
by reading off the coefficients of the second derivatives of the vector field, giving
\begin{equation}
\label{eq:GenVgeffA}
    \hat{g}_{\mu\nu} = z g_{\mu\nu} + 2 z' X_{\mu} X_{\nu} 
    = \frac{2}{\mu^2} \left[ V' g_{\mu\nu} + 2 V'' X_{\mu} X_{\nu} \right]
    \,.
\end{equation}
In conclusion, we obtain the wave equation
\begin{equation}
\label{eq:GenVWavEqA}
    \hat{g}_{\rho\sigma} \nabla^{\rho} \nabla^{\sigma} X_{\mu} - \mathcal{M}_{\mu\rho} X^{\rho} = 0 
\,.
\end{equation}
where we introduced the matrix
\begin{align}
    \mathcal{M}_{\mu\nu} = & z R_{\mu\nu} 
        + z \left(\mu^2 \hat{g}_{\mu\nu} - 3 V'' X_{\mu} X_{\nu}  \right)
        + {\rm l.o.t.}
    \nonumber \\ 
    = &  z R_{\mu\nu} 
        + z \mu^2 \left( \hat{g}_{\mu\nu} - \frac{3 z'}{2}  X_{\mu} X_{\nu}  \right)
        + {\rm l.o.t.}
        \label{eq:MassMatrixA}
\end{align}
and ``l.o.t.'' denotes lower order terms $\sim \nabla X^{\mu}$, that do not contribute if we expand the vector field $X^{\mu}=X^{\mu}_{0}+\epsilon \delta X^{\mu}$ around a constant $X^{\mu}_{0}$.
This tensor can be interpreted as an effective mass matrix for perturbations of the vector field if back-reaction onto the spacetime is negligible.
Note that we expanded the vector field only in this last step: the derivation up to Eq.~\eqref{eq:GenVWavEqA} remains general.

The ghost instability appears if the effective metric satisfies $\hat{g}^{tt}>0$~\cite{Silva:2021jya,Demirboga:2021nrc}.
We identify the condition for the ghost instability by contracting Eq.~\eqref{eq:GenVgeffA} with the time-like normal vector $n^{\mu}$, decomposing the vector field into the scalar potential $\varphi$ and a purely spatial vector $A^{\mu}$ as $X^{\mu} = A^{\mu} + n^{\mu}\varphi$.
We then find
\begin{equation}
\label{eq:GenVGhostConditionA}
\hat{g}_{nn} = \hat{g}_{\mu\nu} n^{\mu} n^{\nu}
    = 2 \varphi^2 z' - z
    = \frac{2}{\mu^2} \left[2 \varphi^2 V'' - V' \right]
\,.
\end{equation}

\section{Supplemental material: 3+1 decomposition and numerical details}
\label{sec-appendix_numerical}

In this appendix we give further details of the background metric, the 3+1 decomposition and the numerical methods used in this work.

\subsection{Metric background}

Following~\cite{East:2017ovw, East:2017mrj, Witek:2012tr}, we use Cartesian Kerr-Schild (KS) coordinates for the fixed background metric
(see also~\cite{Visser:2007fj}). In these coordinates, the spacetime line element is given by:
\begin{eqnarray}
    ds^2 = -  dt^2 +dx^2 + dy^2 + dz^2 + \frac{2Mr^3}{r^4 + a^2z^2} \times \nonumber \\
     \left[ dt + \frac{r(x~ dx + y~ dy)}{a^2 +
    r^2} +   \frac{a(y~ dx - x~ dy)}{a^2 + r^2} +  \frac{(z~dz)}{r}  \right]^2
	\label{KerrKS}
\end{eqnarray}
where $r$ is the Boyer-Lindquist radial coordinate given by the implicit relation
\begin{equation}
	\frac{x^2 + y^2}{r^2 + a^2} + \frac{z^2}{r^2} =  1
	\,,
\end{equation}
and the radial coordinate on the numerical grid is $R = \sqrt{x^2 + y^2 + z^2}$.
Here $a=J/M$ is the Kerr parameter, where $J$ is the angular momentum of the BH, the mass of the BH is $M$, and thus $a/M \in [0,1]$ is the dimensionless spin parameter.  Thus, the BH is entirely parametrized by $J$ and $M$.
We work in a frame in which the angular momentum is aligned with the $z$ direction, without loss of generality.

The metric in the standard  $3+1$ ADM decomposition is given by
\begin{equation} \label{eq-metric}
ds^2=-\alpha^2\,dt^2+\gamma_{ij}(dx^i + \beta^i\,dt)(dx^j + \beta^j\,dt),
\end{equation}
where $\alpha$ denotes the lapse function, $\beta^{i}$ is the shift vector and the 
induced metric is $\gamma_{ij}$.
In Cartesian Kerr-Schild coordinates these are
\begin{equation}
	\alpha = (1 + 2H l^t l^t)^{-1/2},
\end{equation}
\begin{equation}
	\beta^i = - \frac{2H l_t l_i}{1 + 2 H l_t l_t},
\end{equation}
\begin{equation}
	\gamma_{ij}dx^i dx^j = \delta_{ij} + 2H l_i l_j,
\end{equation}
where
\begin{equation}
	H = \frac{Mr^3}{r^4 + a^2 z^2},
\end{equation}
\begin{equation}
	l_\mu = \left( 1, \frac{rx + ay}{r^2 + a^2},  \frac{ry - ax}{r^2 + a^2},
    \frac{z}{r}  \right).
\end{equation}

The extrinsic curvature $K_{ij}$ is derived from its definition as
\begin{equation}
	K_{ij} = \frac{1}{2 \alpha}(D_i\beta_j + D_j \beta_i)
\end{equation}
given that $\partial_t \gamma_{ij} = 0$.

This form of the metric necessitates excision of the singularity, which is achieved by setting the value and time derivative of the field components to zero around the inner horizon. Given sufficient resolution at the outer horizon, the ingoing nature of the metric there prevents the errors this introduces from propagating into the region outside the BH.

The metric is validated by checking that the numerically calculated Hamiltonian and momentum constraints converge to zero with increasing resolution, as do the time derivatives of the metric components, i.e. $\partial_t \gamma_{ij} = \partial_t K_{ij} = 0$ (calculated using the ADM expressions). This ensures that the metric which is implemented is stationary, consistent with it being fixed over the field evolution. 

\subsection{Derivation of 3+1 Proca evolution equations}

Starting from the evolution equation and the constraint in the main text, we derive the Proca evolution equations in a 3+1 decomposition.

It was found crucial to add constraint damping terms to obtain stable numerical simulations of electromagnetic fields~\cite{Palenzuela:2009hx,Hilditch:2013sba,Zilhao:2015tya}.
Following Ref.~\cite{Palenzuela:2009hx,Zilhao:2015tya} we consider the modified field equations
\begin{subequations}
\begin{align}
0 = & \nabla^{\nu} F_{\mu\nu} + \mu^2 X_{\mu} \left(1 + \lambda\, X^{\nu} X_{\nu} \right)
        + \nabla_{\mu} Z - \kappa\, n_{\mu} Z
\,,\\
0 = & \left( 1 + \lambda\, X^{\nu} X_{\nu} \right) \nabla_{\mu} X^{\mu}  
        + \lambda X^{\nu} \nabla_{\nu} \left(X^{\mu} X_{\mu} \right)
        + Z
\,,
\end{align}
\label{eq:EoMsVecLGDamped}
\end{subequations}
where $Z$ denotes an auxiliary (damping) function and  $\kappa$ is the damping parameter.
In practice, we typically set $\kappa = 1.0$.

To perform numerical simulations in full $3+1$ dimensions we 
foliate the spacetime into $3$-dimensional, spatial hypersurfaces characterized by the 
induced metric $\gamma_{ij}$ and the orthonormal vector $n^{\mu}$.
The general line element is determined as in Eq.~\eqref{eq-metric}.
The $3$-metric $\gamma_{ij}$ defines a projection operator
\begin{align}
\gamma^{\mu}{}_{\nu} \equiv & P^{\mu}{}_{\nu} 
        = \delta^{\mu}{}_{\nu} + n^{\mu} n_{\nu}
\,.
\end{align}
We apply the projection operator to decompose the vector field as
\begin{align}
\label{eq:Vector3p1}
X^{\mu} = & A^{\mu} + n^{\mu}\varphi
\,,\quad{\textrm{with}}\quad
A^{\mu} = P^{\mu}{}_{\nu} X^{\nu}
\,,\,\,\,
\varphi = - n_{\nu} X^{\nu}
\,.
\end{align} 
We define the electric and magnetic field as
\begin{align}
\label{eq:DefEB}
E_{i} = & P^{\mu}{}_{i} n^{\nu} F_{\mu\nu}
\,,\quad
B_{i} =   P^{\mu}{}_{i} n^{\nu} \,^{\ast}F_{\mu\nu} 
\,,
\end{align}
where $^{\ast}F_{\mu\nu} = \frac{1}{2}\epsilon_{\mu\nu}{}^{\rho\sigma} F_{\rho\sigma}$ is the dual field tensor,
and $E^{\mu}n_{\mu} = 0 = B^{\mu} n_{\mu}$ by construction.
The Maxwell tensor can now be written as
\begin{align}
\label{eq:FmnInEB}
F_{\mu\nu} = & n_{\mu} E_{\nu} - n_{\nu} E_{\mu} + \epsilon_{\mu\nu\sigma} B^{\sigma}
\,,
\end{align}
where $\epsilon_{\mu\nu\sigma} = n^{\rho} \epsilon_{\rho\mu\nu\sigma}$
is the 3-dimensional Levi-Civita tensor.
Inserting the definition of the Maxwell tensor in terms of the $4$-vector potential, $F_{\mu\nu}=\nabla_{\mu}A_{\nu}-\nabla_{\nu}A_{\mu}$, yields the kinematic relation
\begin{align}
\label{eq:EvoldtAinE}
\dif_{t} A_{i} = & - \varphi D_{i}\alpha - \alpha \left( E_{i} + D_{i} \varphi \right)
\,,
\end{align}
where $\dif_{t} = \p_{t} - \Lie_{\beta}$,
$\Lie_{\beta}$ is the Lie derivative along the shift vector,
and $D_{i}$ is the covariant derivative associated to the induced metric $\gamma_{ij}$.

We decompose the field equations~\eqref{eq:EoMsVecLGDamped}
to find the dynamical evolution equations as in the main text:
\begin{subequations}
\label{eq:EvolVectorAll}
\begin{align}
\label{eq:EvoldtPhi}
\dif_{t} \varphi = & - A^{k} D_{k}\alpha
\\ &
        - \frac{\alpha}{\hat{g}_{\rm nn}} \left(1 + \lambda X^2 \right) \left[ \varphi\,K - D_{k} A^{k} - Z\right]
\nonumber \\ &
        + \frac{2 \lambda \alpha }{\hat{g}_{\rm nn}} [ A^{i} A^{j} D_{i} A_{j} 
\nonumber \\ &        
        \quad \quad \quad - \varphi \left( E^{k} A_{k} - K_{ij} A^{i} A^{j} + 2 A^{k} D_{k}\varphi \right) ]
\,,\nonumber \\
\label{eq:EvoldtE}
\dif_{t} E_{i}   = & 
          D^{k}\alpha \left( D_{i} A_{k} - D_{k} A_{i} \right)
\\ &
        + \alpha \left( D^{k}D_{i}A_{k} - D^{k}D_{k} A_{i} \right)
\nonumber \\ &
        + \alpha \left( K\,E_{i} - 2 K_{ij} E^{j} + D_{i} Z \right)
\nonumber \\ &
        + \mu^2 \alpha A_{i} \left( 1 + \lambda X^2 \right)
~, \nonumber \\
\label{eq:EvolZ}
\dif_{t} Z       = & \alpha\left( D_{i} E^{i} - \kappa Z \right)
+ \alpha\,\mu^2 \varphi \left(1 + \lambda X^2 \right)
\,,
\end{align}
\end{subequations}
where we use the shorthand $X^2 = X^\mu X_\mu = A^i A_i - \varphi^2$, and the coefficient $\hat{g}_{nn}$ is related to the effective metric for the ghost instability defined above.

The constraint 
\begin{align}
C_{E} = & D_{i} E^{i} + \mu^2 \varphi \left(1 + \lambda X^2 \right)
        = 0
\,,
\end{align}
is solved trivially for the initial data at the $t=0$ timeslice, by choosing the form suggested in~\cite{East:2017mrj}:
$A_x = A_0 \gamma^{-1} e^{-\mu r}$, $\phi = E_i = A_y=A_z = 0$.
It is then monitored throughout the evolution to verify that it is satisfied, and we use it to check for convergence in the simulations, as detailed in the following section.

\begin{figure}[]
\includegraphics[width=0.9\columnwidth]{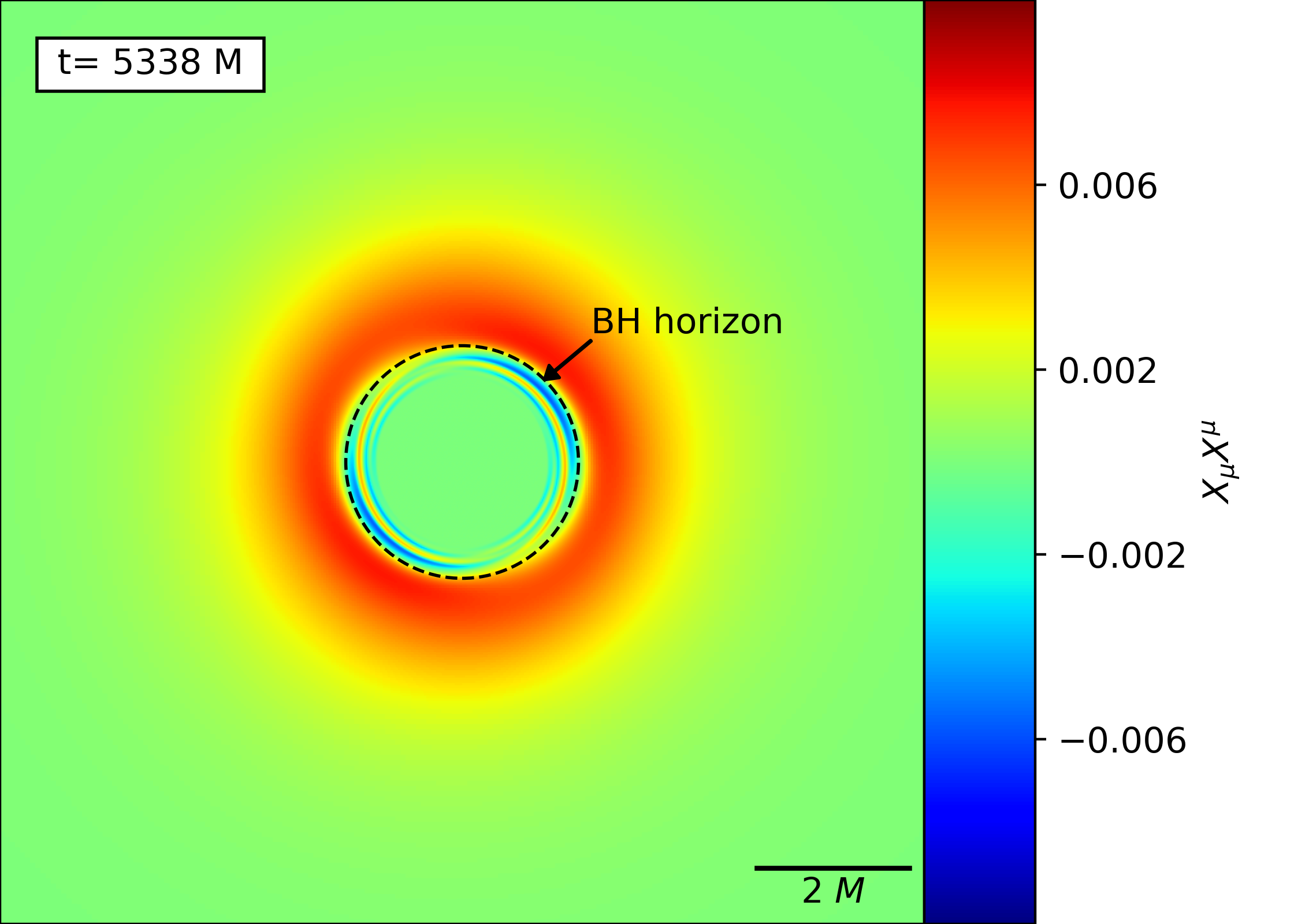}
\caption{
Slice through the center of the BH in the plane perpendicular to the spin axis, showing the value of $X^\mu X_\mu$ for the superradiant state that is obtained during the evolution. There are regions in which $X^\mu X_\mu$ is negative as well as positive. In the slice shown the regions of negative $X^2$ are within the horizon, but this is not the case throughout the simulation. At some points the negative $X^2$ regions do reside outside the horizon, and thus have a causal effect on the evolution.
\label{fig-Xsquared}
}
\end{figure}

\begin{figure}[]
\includegraphics[width=0.9\columnwidth]{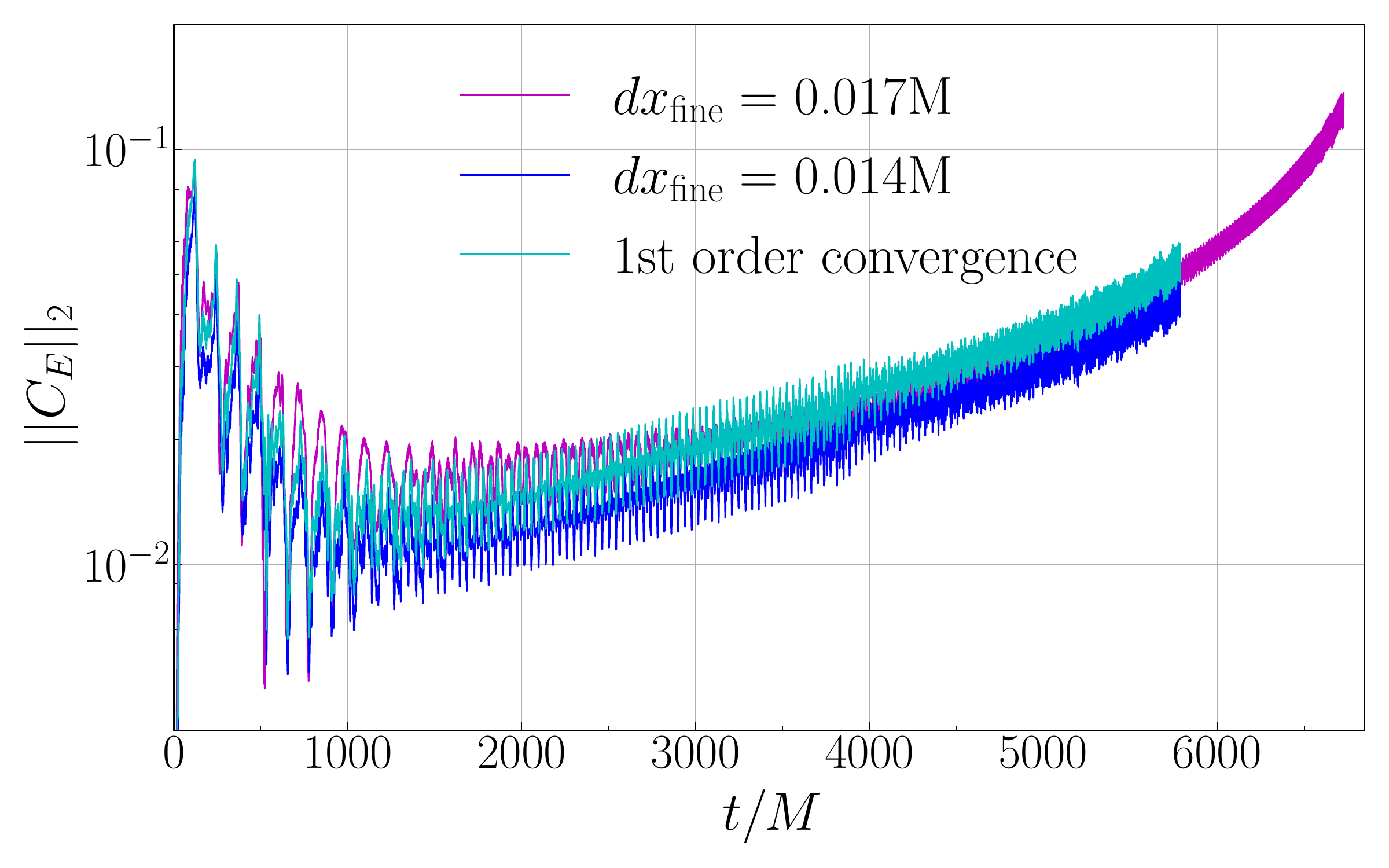}
\caption{
Convergence test for the case where $\lambda = -1$. The norm of the constraint violation across the grid reduces to zero at approximately first/second order when comparing two different resolutions at the BH horizon. Note that since it has a slightly faster growth rate, the higher-resolution simulation reaches the instability and ends before the low resolution one. Note that the timing of the breakdown still coincides with the effective metric changing signature: simply the growth of the cloud amplitude is slightly faster with the better resolution, so the failure point is reached earlier in simulation time.
\label{fig-SR_convergence}
}
\end{figure}

\begin{figure}[]
\includegraphics[width=0.5\textwidth]{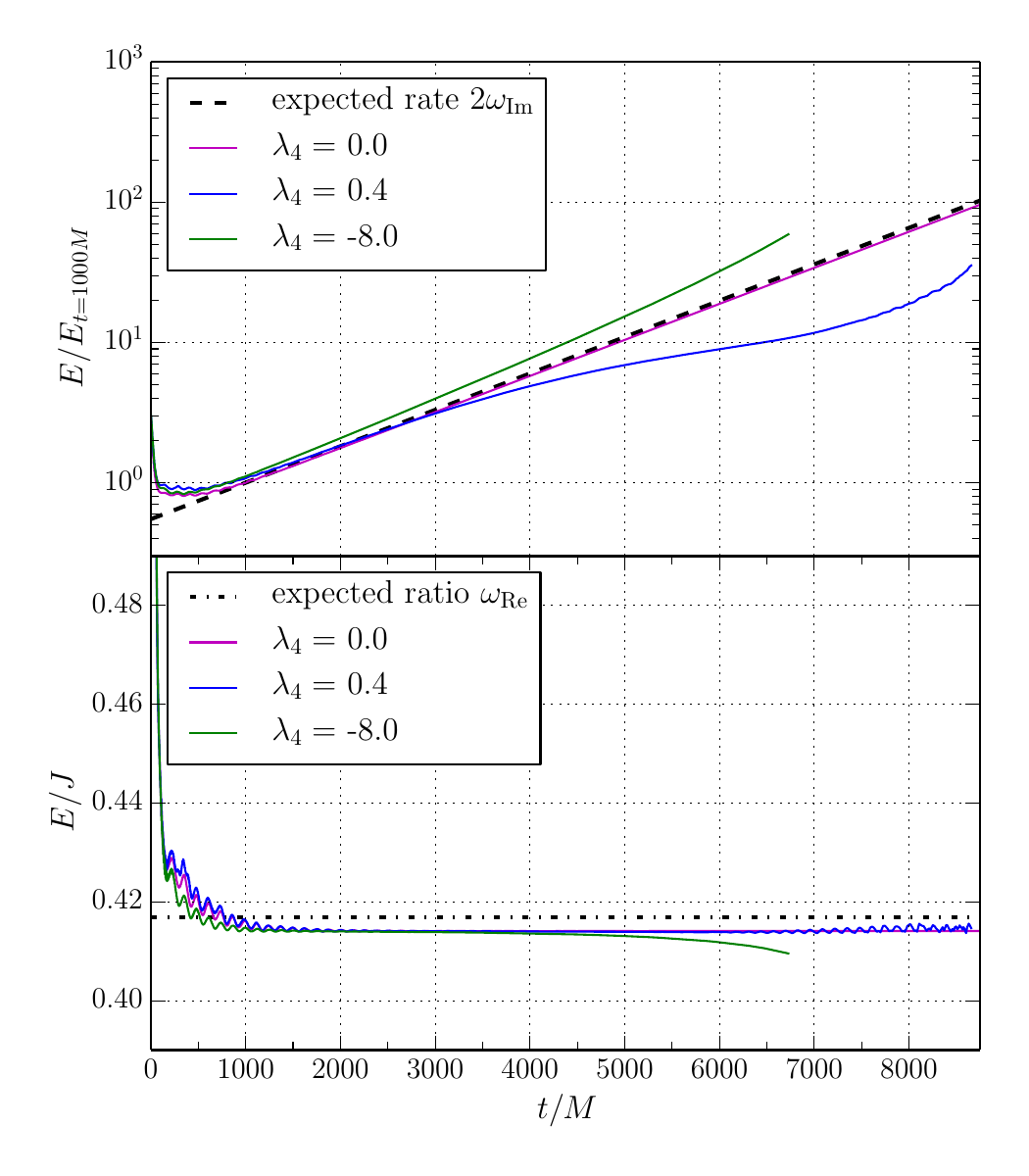}
\caption{
Growth of the total energy in the superradiant cloud, normalized by its value at $t/M=1000$, and its frequency as measured by $\omega_{\rm Re} = E/J$, compared to the results obtained numerically and semi-analytically in previous works~\cite{East:2017mrj, Dolan:2018dqv}. We are able to recover the correct rate in the purely massive case, and initially in the self-interacting cases, which diverge at later times as expected, once the self-interactions become large.
\label{fig-SR_rates}
}
\end{figure}

\subsection{Numerical details and convergence}

We use an adapted version of the {\sc GRChombo} numerical relativity framework
\cite{Clough:2015sqa,Andrade2021,Radia:2021smk} with the metric components and their derivatives
calculated analytically at each point rather than stored on the grid. The
evolution of the Proca field components follows the standard method of lines,
with a fourth-order Runge-Kutta time integrator and fourth-order finite difference stencils. 
An image of the resulting superradiant profile for $X^2$ in the plane perpendicular to the spin axis is shown in Fig.~\ref{fig-Xsquared}.
We note that the oscillations seen in Fig. 2 of the main text arise from mixing of the superradiant modes, which happens once self-interactions become significant. This leads to different frequency content as higher order modes are excited. The oscillations are quite well resolved once the x-axis is expanded, but there are occasional jumps that come from the fact that we are not following a single point, but instead taking the minimum value of $\hat{g}_{nn}$ across the entire spatial slice: the position with the smallest value can change over time, which makes its evolution look less smooth. If one looks instead at the value of the evolved fields or $X^\mu X_\mu$, the profile and time evolution appear smooth up until the breakdown.

We use a box of full width $L=410M$ with $N=192$ grid points across it and 7 levels of 2:1 refinement, resulting in a finest grid spacing of $dx_{\rm fine} = 0.017M$. We repeat the calculation at a finer resolution with $N=224$ and $dx_{\rm fine} = 0.014M$ to check convergence.

For example, as illustrated in Fig.~\ref{fig-SR_convergence}, we check the convergence of the constraint to zero, and find that it converges at between first and second order. This is lower than the 3rd/4th order expected from the evolution and refinement errors, partly due to the constraint being summed only over the region outside the horizon. Due to the excision region being a ``lego sphere'', the actual physical volume over which the constraint is integrated to calculate the 2-norm inevitably depends on the resolution. Since the constraint is largest near the horizon, this generally results in a higher amount of constraint violation in the finest resolution case as the inner sphere is better resolved. Nevertheless, the constraint does reduce as the resolution increases, and we find consistent results at all resolutions.

We also checked that, following an initial transient period, all cases grow smoothly at the expected superradiant rate, with the frequencies obtained in~\cite{East:2017mrj, Dolan:2018dqv}, as shown in Fig.~\ref{fig-SR_rates}, and that the fluxes of the Proca field across the horizon and out of the domain reconcile to the growth of the cloud mass and angular momentum.

\bibliography{SelfIntSR}

\end{document}